# Diquark approach to calculating the mass and stability of multiquark states

A. R. Haghpayma*

Department of Physics, Ferdowsi University of Mashhad, Mashhad, Iran

**Abstract.** Diquarks may play an important role in hadronic physics particularly near the phase transitions ( chiral, deconfinement points ), current lattice QCD determinations of baryon charge distributions do not support the concept of substantial u - d scalar diquark clustering as an appropriate description of the internal structure of nucleon, thus vector diquarks are more favourable. By using of vector diquark ideas in the chiral limit diquark correlations in the relativistic region and imposing HF interactions between quarks in a vector diquark we calculated the mass of two multiquark states ( $\Theta^+$ pentaquark and H dibaryon ), also by using of tunneling method we simultaneously calculated their decay width .Our approach is general and extendable to all multiquark states. Although the new JLab experiment shows that there is no $\Theta^+$ in the mass region 1480 Mev $< m_{\Theta^+} <$1700 MeV, our model predicts that for $m_{\Theta^+} >$1700 MeV its width would be $\Gamma_{\Theta^+} > 53$ Mev that shows there is a possibility for the existence of multiquark states which are too wide to be detected by experiments. The paper achieves $M_{\Theta^+} \simeq 1540$ MeV and $\Gamma_{\Theta^+} \simeq 1.30$ MeV for the mass and width of $\Theta^+$ which are in good agreement with experimental limits.

## 1 Introduction

QCD is believed to be the underlying theory of the strong interaction which has three fundamental properties : asymptotic freedom, colour confinement, approximate chiral symmetry and its spontaneous breaking and in high energy level QCD has been tested up to the 1% level.

The behaviour of QCD in the low energy is nonperturbative and the $SU_c(3)$ colour group structure is non- abelian. However,beside the conventional mesons and baryons with the $q\bar{q}$ and 3q quark structure respectively, QCD itself does not exclude the existence of the nonconventional states such as glueballs ( gg , ggg , ...... ) hybrid mesons ( $q\bar{q}$ g ) and other multi - quark states ( qqqq , qqqqq )[1].

Do other multiquark hadrons exist? 4q , 6q ,7q , ... , Nq, is there an upper limit for N ? study of these issues will deepen our understanding of the low- energy sector of QCD.

Thus we need to understand the underlying dynamics of these states and concepts such as confined quarks and gluons or fundamental concepts such as spin and mass of a confined quark and a free quark or the decay width of a hadron. There is a basic question

Can low - energy QCD describe the underlying dynamical forces between quarks and gluons in multiquark states and generate their mass and width correctly ?

It is very difficult to calculate the whole hadron spectrum from first principles in QCD[2], under such a circumstance various models which are QCD - based or incorporate some important properties of QCD were proposed to explain the hadron spectrum and other low-energy properties. For more details, see Ref [3] in which the papers are explained further to some extent.

In the high temperature case, it is evident that strong interacting matter exists in the form of a quark - gluon plasma, but at low temperatures where the baryon density is high, bosonic states, diquarks and even longer quark clusters may play an important role, at this region of energy the diquarks form a Bose condensate and also there is a possibility of a pentaquark.At 2 Gev/200 MeV hadronic systems can undergo a phase transition from the hadronic to a deconfined state of quarks and gluons, quark - gluon plasma (QGP) where quarks behave like free fermi gas. Thus diquarks may play an important role in hadronic physics particularly near the phase transitions ( chiral, deconfinement points ).

In fact the notion of diquark is as old as the quark model itself, for example Gell - Mann mentioned the possibility of diquarks in his original paper on quarks. From the description of baryons as composed of a constituent diquark and quark by Ida and Kobayashi[4] and Lichtenberg and Tassie[5] up to now many articles have been written on this subject[6,7].

The microscopic origin of diquarks is not completely clear and its connection with the fundamental theory (QCD) not fully understood, apart from the cold asymptotic high dense baryon case, in which the quarks form a fermi surface and perturbative gluon interactions support the existence of a diquark BCS state[8] known as colour superconductor.

There is hyperfine interaction (HF) between quarks in a diquark which indeed is induced by instantons, and besides of these interactions there is confinement interaction between them. The hyperfine energy lies between the deconfinment and the asymptotic freedom regions.

There are several models in which there are scalar and/or tensor diquarks and successfully describe hadron properties, but current lattice QCD (LQCD) determinations of baryon charge distributions do not support the concept of substantial u - d scalar diquark clustering as an appropriate description of the internal structure of the baryons. Furthermore, the overlap of a displaced scalar diquark source and the positive parity ground state for $\Theta^+$ pentaquark is 0.03,[9] indicating that vector diquarks maybe more favourable.Both the LQCD ad QCD sum rule results reject existence of a low-lying positive parity state for $\Theta^+$[10].this paper calculate the mass and decay width of two interesting multiquark states by using of vector diquarks.

## 2 Diquark approach

Effective field theories are not just models, they represent very general principles such as analyticity, unitarity, cluster decomposition of quantum field theory and the symmetries of the systems. For example chiral perturbation theory ($\chi$ PT ) describes the low - energy behaviour of QCD within the framework of an effective theory(at least in the meson sector), for a recent review see e.g. [11] , and for some examples [12] .

With a conventional $\chi$PT if we consider group theoretical clustering between quarks[13] and hyperfine QCD interactions between them, we have an correlated $\chi$PT between quarks or correlated perturbative chiral quark model CP$\chi$QM. In fact this theory describes correctly the physics of QCD[14] between the confinement scale and the chiral symmetry breaking scale. In other case we are working with non - correlated theories.[15]

It is well known that the multiquarks cannot be a simple n-quark states in ground state, because they would freely recombine and decay in baryons and mesons, with a very broad decay width.

If we discuss about an correlated P$\chi$QM with explicit symmetry on configuration, the hyperfin interactions ( FS , CS ) would be considered on the n-quark subsystems( diquarks,...), and a special form of confining potential.

Although the diquark approach in our model is extendable to all possible multiquark states, we consider the $\Theta^+$ pentaquark and H dibaryon as two examples of multiquark states and use the diquark approach for calculating their mass and decay width simultaneously. The limit on the $\Theta^+$ width is very low[16] $\Gamma_\Theta \simeq 1$ $MeV << \Gamma_B$ for conventional baryon resonances and in quark model this picture is less clear.

## 3 Pentaquark states

Here, pentaquark states are studied in a correlated perturbative chiral quark model (CP$\chi$QM) Hamiltonian.Any pentaquark state can be formally decomposed in combinations of simpler coulour-singlet clusters, the energy of the state is computed with the state matrix element of the Hamiltonian.

The paper suggest that the $\Theta^+$ baryon is a composed state of two vector diquarks and a single antiquark, the spatially wave function of these diquarks has a p - wave and a s - wave in angular momentum in the first and second versions of our model respectively.

*e - mail: haghpeima@wali.um.ac.ir

The $[\,2\,]^{f}$ flavour symmetry of each diquark leads to $[\,22\,]^{f}_{\bar{6}}$ flavour symmetry for $q^4$ and the $[\,2\,]^{s}$ spin symmetry of each diquark leads to $[\,22\,]^{s}$ and $[31]^{s}$ spin symmetry for $q^4$ in the first version and second version of the model respectively.

The colour symmetry of each diquark is $[\,11\,]^{c}$ and for the first version of the model we assume $[\,2\,]^{c}$ for one of the diquark pairs, this leads to $[\,211]^{c}$ colour symmetry for $q^4$.

The orbital symmetry of each diquark is $[\,2\,]^{o}$ and for the first version we assume $[\,11\,]^{o}$ for one of the diquark pairs, this leads to $[\,31\,]^{o}$ and $[\,4\,]^{o}$ orbital symmetry for $q^4$ in the first and second version of the model respectively.

As the result of these considerations we would have for $q^4$ contribution of quarks ,( $[\,1111\,]^{oc}$ and $[\,4\,]^{fs}$ ) and ( $[\,211\,]^{oc}$, $[\,31\,]^{fs}$ ) for the first and second versions ; and $[\,1111\,]^{ocfs}$ for $q^4$ is the same for the two versions and lead to a totally antisymmetric wave function for $q^4$ due to pauli principle.

By introducing the quark and antiquark operators and by taking direct product of two diquarks and one antiquark the $\Theta^+$ state can be represented by $SU_f(\,3\,)$ tensors.

We denote a quark with $q_i$ and antiquark with $q^i$ in which $i$=1,2,3 denote u , d , s and impose normalization relations as

$$(q_i, q_j) = \delta_{ij},\ (q^i, q^j) = \delta^{ij},\ (q_i, q_j) = 0$$

and in $(p\ ,\ q)$ notation we consider a tensor $T^{b_1,\ldots,b_q}_{a_1,\ldots,a_p}$ which is completely symmetric in upper and lower indices and traceless on every pair of indices.

We introduce $S_{jk} = \frac{1}{\sqrt{2}}(q_j q_k + q_k q_j)$ and $A_{jk} = \frac{1}{\sqrt{2}}(q_j q_k - q_k q_j)$ ,then for a quark and an antiquark we would have

quark $\quad 3 : \quad T_i \qquad ijk = 1,2,3$

antiquark $\quad \bar{3} : \quad T^i = \epsilon^{ijk} A_{jk},$

where $\epsilon^{ijk}$ is the levi-civita tensor and we have

$$(T^i, T^j) = 4\delta^{ij}.\ (T_i, T_j) = 4\delta_{ij}.\ (S_{jk}, S_{lm}) = \delta_{jl}\delta_{km} + \delta_{jm}\delta_{kl}, \quad (S_{jk}, T^i) = 0$$

for $q^4$ we have

$$(\mathbf{3}\otimes\mathbf{3}\otimes\mathbf{3}\otimes\mathbf{3}) = (\mathbf{6}\oplus\bar{\mathbf{3}}) \otimes (\mathbf{6}\oplus\bar{\mathbf{3}}) = (\mathbf{6}\otimes\mathbf{6}) \oplus (\mathbf{6}\otimes\bar{\mathbf{3}}) \oplus (\bar{\mathbf{3}}\otimes\mathbf{6}) \oplus (\bar{\mathbf{3}}\otimes\bar{\mathbf{3}}) = (\mathbf{15}_1 \oplus \mathbf{15}_2 \oplus \bar{\mathbf{6}}) \oplus (\mathbf{15}_2 \oplus \mathbf{3}) \oplus (\mathbf{15}_1 \oplus \mathbf{3}) \oplus (\bar{\mathbf{6}} \oplus \mathbf{3}) \quad (1)$$

and for $q^4\bar{q}$ we have

$$[\,\mathbf{3}\otimes\mathbf{3}\otimes\mathbf{3}\otimes\mathbf{3}] \otimes \bar{\mathbf{3}} = (\,[\,\mathbf{15}_1 \oplus \mathbf{15}_2 \oplus \bar{\mathbf{6}}\,] \oplus [\mathbf{15}_2 \oplus \mathbf{3}] \oplus [\mathbf{15}_1 \oplus \mathbf{3}] \oplus [\bar{\mathbf{6}} \oplus \mathbf{3}\,]\,) \otimes \bar{\mathbf{3}}$$
$$= 2(\,\mathbf{15}_1 \otimes \bar{\mathbf{3}}) \oplus 2(\,\mathbf{15}_2 \otimes \bar{\mathbf{3}}) \oplus 2(\bar{\mathbf{6}} \otimes \bar{\mathbf{3}}\,) \oplus 2(\mathbf{3} \otimes \bar{\mathbf{3}}\,)$$
$$= 2\,(\,\mathbf{35} \oplus \mathbf{10}) \oplus 2(\mathbf{27} \oplus \mathbf{10} \oplus \mathbf{8}\,) \oplus 2(\,\overline{\mathbf{10}} \oplus \mathbf{8}) \oplus 2(\mathbf{8} \oplus \mathbf{1}\,) \quad (2)$$

The tensor notation of all of these multiplets can be constructed as $T_i$ and $T.^i$

For example the tensor notations for $\overline{10}$ and 8 are

$$\overline{10}:\quad T^{ijk} = \frac{c_1}{\sqrt{3}}\left(S^{ij}\bar{q}^k + S^{jk}\bar{q}^i + S^{ki}\bar{q}^j\right) + \frac{c_2}{\sqrt{3}}\left(T^{ij}\bar{q}^k + T^{jk}\bar{q}^i + T^{ki}\bar{q}^j\right), \quad (3)$$

$$8:\quad P^i_j = \frac{c_1}{\sqrt{2}}\left(T_j\bar{q}^i - \frac{1}{3}\delta^i_j T_m\bar{q}^m\right) + \frac{c_2}{\sqrt{2}}\left(Q_j\bar{q}^i - \frac{1}{3}\delta^i_j Q_m\bar{q}^m\right) + \frac{c_3}{\sqrt{2}}\left(\widetilde{Q}_j\bar{q}^i - \frac{1}{3}\delta^i_j \widetilde{Q}_m\bar{q}^m\right.$$
$$+ \frac{c_4}{\sqrt{3}}\epsilon_{jab}S^{ia}\bar{q}^b + \frac{c_5}{\sqrt{3}}\epsilon_{jab}T^{ia}\bar{q}^b + \frac{c_6}{\sqrt{15}}T^i_{jk}\bar{q}^k + \frac{c_7}{\sqrt{15}}\widetilde{T}^i_{jk}\bar{q}^k + \frac{c_8}{\sqrt{15}}S^i_{jk}\bar{q}^k. \quad (4)$$

in which

$$\widetilde{Q}_m = \frac{1}{\sqrt{8}}T^l S_{ml}, \qquad \widetilde{T}^i_{jk} = T^i S_{jk} - \frac{1}{\sqrt{2}}\left(\delta^i_j\delta^m_k + \delta^i_k\delta^m_j\right)\widetilde{Q}_m,$$
$$T^{ij} = \frac{1}{\sqrt{6}}\epsilon^{iab}\epsilon^{jcd}S_{ac}S_{bd}, \qquad \widetilde{Q}_m = \frac{1}{\sqrt{8}}T^l S_{ml},$$
$$S^i_{jk} = \frac{}{\sqrt{2}}\epsilon^{ilm}\left(S_{jl}S_{km} + S_{kl}S_{jm}\right). \qquad T^i_{jk} = S_{jk}T^i - \frac{1}{\sqrt{2}}\left(\delta^i_j\delta^m_k + \delta^i_k\delta^m_j\right)Q_m.$$

The nomenclature for pentaquark states based on hypercharge is as follow

$Y = 2 \quad \Theta, \qquad Y = 0 \quad \Sigma, \Lambda, \qquad Y = -2 \quad \Omega.$
$Y = 1 \quad N, \Delta, \qquad Y = -1 \quad \Xi\ , \qquad Y = -3 \quad X.$

which in the notation $T^{b_1,\ldots,b_q}_{a_1,\ldots,a_p}$ we have

$$Y = p_1 - q_1 + p_2 - q_2 + \frac{3}{2}(p - q),$$
$$I_3 = \frac{1}{2}(p_1 - q_1) - \frac{1}{2}(p_2 - q_2), \qquad p_1 + p_2 + p_3 = p \text{ and } q_1 + q_2 + q_3 = q$$
$$Q = I_3 + Y/2\ .$$

Now the $SU_f(\,3\,)$ symmetry lagrangian would be

$$\mathcal{L}_{\overline{10}-8_3} = g_{\overline{10}-8_3}\,\epsilon^{ilm}\overline{T}_{ijk}B^j_l M^k_m + (\text{H.c.}) = -\sqrt{6}\ \overline{\Theta}^+\left(pK^0 - nK^+\right) + \ldots.. \quad (5)$$

where $B^j_l$ and $M^k_m$ are the 3 - quark baryon octet and meson octet respectively and $g_{\overline{10}-8_3}$ is the universal coupling constant.

The flavour symmetry contribution for $q^4$ configuration comes from the decomposition formula (2) for $q^4\bar{q}$ in which we have two $\bar{6}_f$ the second one is used in J W model[13] and the first one is used in our model due to vector diquark contribution of it.

Thus the tensor notations for $\overline{10}_f$ in our model comes from Eq (3) with ($C_1$=0 and $C_2 \neq 0$ ) also the tensor notations for $8_f$ in our model comes from Eq (4) with $C_i = 0$ $i \neq 5$ ,$C_5 \neq 0$ .

Inserting $\overline{10}_f$ tensor notation into $\mathcal{L}_{\overline{10}-8_3}$ symmetry lagrangian leads to $SU_f(\,\bar{3}\,)$ symmetry interactions which experimental evidence of them would be explored and the discovery of them will support the vector diquark approach.

Considering flavour configuration of $q^4$ in Eq(1), one can see that there are $15_1$ and $15_2$ multiplets which coms from $6 \otimes 6$ normal products and this leads to two 45 multiplets for flavour configurations of $q^4\bar{q}$ in Eq(2) in which there is octet, decuplet , 27 plet and 35 plet.

The $SU_f(\,3\,)$ configurations of these multiplets are $[\,321\,]_8$, $[\,411\,]_{10}$ $[\,42\,]_{27}$ and $[\,51\,]_{35}$ and the $SU_{fs}(\,6\,)$ configurations of them are

$$[\,33111]_{560},\ [\,321111]_{70},\ [\,41111]_{56} \text{ and } [\,51111]_{700}$$

The tensor notations of $15_1$ and $15_2$ are $T_{jklm}$ and $S^i_{jk}$ respectively

$$T_{jklm} = \frac{1}{\sqrt{6}}\left(S_{jk}S_{lm} + S_{lk}S_{jm} + S_{jm}S_{kl} + S_{lj}S_{km} + S_{km}S_{jl} + S_{lm}S_{jk}\right),$$
$$S^i_{jk} = \frac{1}{\sqrt{2}}\epsilon^{ilm}\left(S_{jl}S_{km} + S_{kl}S_{jm}\right) \quad (6)$$

and by multiplying to tensor notation of $\bar{q}$ ( $T^i$ ) we lead to tensor notations of ( 8 , 10 , 27 , 35 ) plets.

In fact there is noting in quark model to prevent us for constructing such multiplets in which they have vector diquarks and the discovery of them will be evidence supporting the vector diquark approach.

Briefly the spin - flavour - color and parity of our model for the first version and second one are as follows

$$\left|\{\boldsymbol{QQ}\}^{\ell=1,\mathbf{3}_c,\bar{\mathbf{6}}_f}\ \bar{q}^{\ j=\frac{1}{2},\bar{\mathbf{3}}_c,\bar{\mathbf{3}}_f}\right\rangle^{J^\Pi=(\frac{1}{2}^+\oplus\frac{3}{2}^+)\ ,\mathbf{1}_c,(\overline{\mathbf{10}}_f\oplus\mathbf{8}_f)} \quad (7)$$

$$\left|\{\boldsymbol{QQ}\}^{\ell=0,\mathbf{3}_c,\bar{\mathbf{6}}_f}\ \bar{q}^{\ j=\frac{1}{2},\bar{\mathbf{3}}_c,\bar{\mathbf{3}}_f}\right\rangle^{J^\Pi=(\frac{1}{2}^-\oplus\frac{3}{2}^-),\mathbf{1}_c,(\mathbf{8}_f\oplus\overline{\mathbf{10}}_f)} \quad (8)$$

We have considered $[\,4\,]^{fs}_{126}$ and $[\,31\,]^{fs}_{210}$ for the flavour - spin configurations of $q^4$ in the first and second versions of our model respectively, this leads to $[\,51111]_{700}$ and $[\,42111]_{1134}$ for the flavour - spin configurations of $q^4\bar{q}$ , but if one assume the angular momentum $\ell = 1$ for the four quarks in $q^4$ there are several allowed $SU_{fs}(\,6\,)$ representations for $q^4\bar{q}$ which are $[\,51111\,]_{700}$, $[\,42111\,]_{1134}$, $[33111\,]_{560}$ and $[\,32211\,]_{540}$ based on $[\,4\,]^{fs}$, $[\,31\,]^{fs}$, $[\,22\,]^{fs}$ and $[\,211\,]^{fs}$ $SU_{fs}(\,6\,)$ representations for $q^4$ respectively.

The constituent quark model have not been yet derived from QCD, therefore it is useful to consider the Effective Hamiltonian approach and by using of diquark ideas in the chiral limit diquark correlations in the relativistic region and imposing HF interactions between quarks in a diquark,fig.1,we led to introducing a conventional Hamiltonian for pentaquark

$$H(q^4\bar{q}) = T(q^4\bar{q}) + \overset{\text{bin}}{V}(q^4\bar{q}) + \overset{\text{cs}}{V}(q^4\bar{q}) \quad (9)$$

containing Kinetic energy, binding and hyperfine interactions

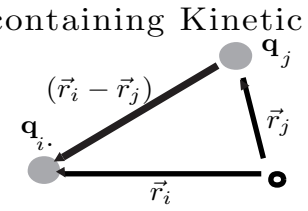

$$\overset{\text{bin}}{V}(q_i q_j) = a\frac{\vec{\lambda}_i}{2}\cdot\frac{\vec{\lambda}_j}{2}(\vec{r}_i - \vec{r}_j)^2. \quad (10)$$

$$\overset{\text{cs}}{V}(q_i q_j) = -\mathrm{b}\frac{\vec{\lambda}_i}{2}\cdot\frac{\vec{\lambda}_j}{2}\vec{S}_i.\vec{S}_j. \quad (11)$$

FIG.1 :Diquark configuration in coordinate space.

where $a$ and b are iteraction constantes also $\lambda_i$ and $S_i$ are colour and spin matrices respectively.

We have neglected confinement and flavour - spin interactions and considered $V^{cs}$ as a noncontact interaction.

Now in the $q^4\bar{q}$ rest frame, we define the internal variables,fig.2

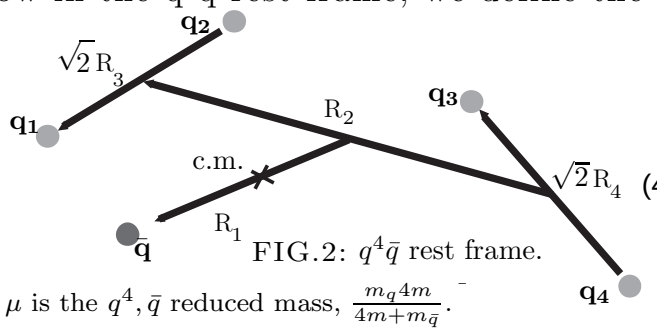

FIG.2: $q^4\bar{q}$ rest frame.

$\mu$ is the $q^4, \bar{q}$ reduced mass, $\frac{m_q 4m}{4m + m_{\bar{q}}}$.

$$\vec{r}_1 = -\frac{\mu}{4m}\vec{R}_1 + \frac{1}{2}\vec{R}_2 + \frac{1}{\sqrt{2}}\vec{R}_3$$
$$\vec{r}_2 = -\frac{\mu}{4m}\vec{R}_1 + \frac{1}{2}\vec{R}_2 - \frac{1}{\sqrt{2}}\vec{R}_3$$
$$\vec{r}_3 = -\frac{\mu}{4m}\vec{R}_1 - \frac{1}{2}\vec{R}_2 + \frac{1}{\sqrt{2}}\vec{R}_4 \quad (47)$$
$$\vec{r}_4 = -\frac{\mu}{4m}\vec{R}_1 - \frac{1}{2}\vec{R}_2 - \frac{1}{\sqrt{2}}\vec{R}_4$$

Thus we would have for the Kinetic energy $\vec{r}_{\bar{q}} = \frac{\mu}{m_{\bar{q}}}\vec{R}_1$

$$T(q^4\bar{q}) = \frac{\vec{\nabla}^2_{R_1}}{2\mu} + \frac{\vec{\nabla}^2_{R_2}}{2m} + \frac{\vec{\nabla}^2_{R_3}}{2m} + \frac{\vec{\nabla}^2_{R_4}}{2m} \quad (12)$$

The orbital wave function is

$$\psi_m = N[a_{R_2}Y_{1m}(\hat{R}_2)e^{-a^2 R_2^2/2}][Y_{00}(\hat{R}_3)e^{-\beta^2 R_3^2/2}][Y_{00}(\hat{R}_4)e^{-\beta^2 R_4^2/2}][Y_{00}(\hat{R}_1)e^{-\gamma^2 R_1^2/2}] \quad (13)$$

we have

$$\langle \overset{\text{bin}}{V}{}^{q^2\text{-}q^2}(q^4)\rangle_4 = \frac{5}{3}a(R_3^2 + R_4^2) + \frac{2}{3}aR_2^2 \qquad \langle R_1^2\rangle = \frac{3}{2\gamma^2}.\quad \langle R_2^2\rangle = \frac{5}{2a^2} \quad \langle R_3^2\rangle = \langle R_4^2\rangle = \frac{3}{2\beta^2}$$
$$\langle \overset{\text{bin}}{V}{}^{\bar{q}\text{-}q}(\bar{q})\rangle = \frac{1}{3}a(R_3^2 + R_4^2) + \frac{1}{3}aR_2^2 + \frac{4}{3}aR_1^2 \quad (14)$$
$$\langle \overset{\text{cs}}{V}{}^{q^2\text{-}q^2}(q^4)\rangle = \frac{2}{3}\mathrm{b}(\vec{S}_1.\vec{S}_2 + \vec{S}_3.\vec{S}_4) \qquad \langle\vec{\nabla}^2_{R_1}\rangle = \frac{3\gamma^2}{2}.\quad \langle\vec{\nabla}^2_{R_2}\rangle = \frac{5a^2}{2}.\quad \langle\vec{\nabla}^2_{R_3}\rangle = \langle\vec{\nabla}^2_{R_4}\rangle = \frac{3\beta^2}{2}$$
$$+ \frac{1}{6}\mathrm{b}(\vec{S}_1.\vec{S}_3 + \vec{S}_1.\vec{S}_4 + \vec{S}_2.\vec{S}_3 + \vec{S}_2.\vec{S}_4)$$
$$\langle \overset{\text{cs}}{V}{}^{\bar{q}\text{-}q^4}(\bar{q})\rangle = \frac{1}{3}\mathrm{b}\vec{S}(q^4).\vec{S}(\bar{q})$$

Now we consider this solution for the second version of our model which $\ell = 0$, after calculating the mass of a vector diquark from colour-spin interaction between quarks $M_{ud} \simeq$ 520Mev and assuming $m_s \simeq$ 450 MeV and T$\simeq$ 50MeV the resulted mass of $\Theta^+$ pentaquark would be 1540 MeV.

Since the diquark masses ( e.x, scalar or vector, .... ) are smaller than the constituents, they are stable against decay near mass shell, in such a configuration, the diquarks are nearby and tunneling of one of the quarks between the two diquarks may take place.

We suppose that the small Decay widths of $\Theta^+$ is due to tunneling of one of the quarks between the two vector diquarks. Thus in the decay process $\Theta^+ \to K^+N$ a (d) quark tunnels from a diquark (ud) to the other diquark to form a nucleon (udd) and an off-shell (u) quark which is annihilated by the anti- strange quark. The decay width of this process is

$$\Gamma_{\Theta^+} \simeq 5.0\, e^{-2S_0} \frac{g^2 g_A^2}{8\pi f_K^2} |\psi(0)|^2 . \qquad (15)$$

Which we have used WKB approximation for the tunneling amplitude and $\Delta E = (m_u + m_d - M_{ud})$ .The $\psi(0)$ is the 1S wave function of quark - diquark at the origin and can be written as

$$\psi(0) = \frac{2}{a_0^{3/2}} \frac{1}{\sqrt{4\pi}}, \qquad (16)$$

According to our model the Kinetic energy $T \simeq \frac{\vec{\nabla}_{R_2}^2}{2m} = \frac{3a^2}{4m} \simeq 50$ MeV this leads to $a \simeq 148$ and then

$$r_0 = \langle R_2 \rangle = \sqrt{\frac{5}{2a^2}} \simeq 0.010 \text{ MeV}^{-1} \qquad (17)$$

where $g^2 = 3/03$, $g_A = 0.75$ from the quark model and $\Delta E = (m_u + m_d) - M_{ud}$ and $M_{ud} \simeq 520$ MeV in our model.

Inserting this values into Eq (15) we find

$$\Gamma_{\Theta^+} \simeq 1.30 \text{ MeV} \qquad (18)$$

Which is unusually narrow and camparable with the experimental limit[16,17] $\Gamma_{\Theta^+} \simeq 1$ MeV.

We have calculated the $\Theta^+$ pentaquark decay width for a range of its mass, 1500 Mev $< M_{\Theta^+} <$ 1700 MeV. we find $\Gamma(\Theta^+ \to K^+ n) \simeq 0.0003 \sim 53.670$ MeV, Table1 and fig.3.

| $M_{\Theta^+}$ MeV | $T$ MeV | $r_0$ MeV$^{-1}$ | $\Gamma_{\Theta^+}$ MeV |
|---|---|---|---|
| 1500 | 10 | 0.023 | 0.0003 |
| 1510 | 20 | 0.016 | 0.0700 |
| 1520 | 30 | 0.013 | 0.4800 |
| 1530 | 40 | 0.012 | 0.9200 |
| 1540 | 50 | 0.011 | 1.3000 |
| 1550 | 60 | 0.009 | 3.8300 |
| 1560 | 70 | 0.008 | 5.1200 |
| 1570 | 80 | 0.008 | 5.1200 |
| 1580 | 90 | 0.007 | 13.590 |
| 1590 | 100 | 0.007 | 13.590 |
| 1600 | 110 | 0.007 | 13.590 |
| 1610 | 120 | 0.006 | 17.760 |
| 1620 | 130 | 0.006 | 17.760 |
| 1630 | 140 | 0.006 | 40.890 |
| 1640 | 150 | 0.006 | 40.890 |
| 1650 | 160 | 0.005 | 53.670 |
| 1660 | 170 | 0.005 | 53.670 |
| 1670 | 180 | 0.005 | 53.670 |
| 1680 | 190 | 0.005 | 53.670 |
| 1690 | 200 | 0.005 | 53.670 |
| 1700 | 210 | 0.005 | 53.670 |

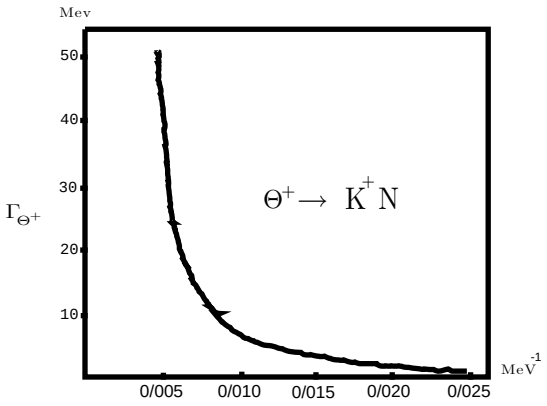

FIG.3:The calculated width of the negative- parity $\Theta^+$ pentaquark is shown for the decay $\Theta^+ \to K^+N$

Table.1: pentauark $\Theta^+$ decay width for a range of its mass. $r_0$ is the distance between two vector (ud) diquarks.

Now the paper extend the diquark model which previously formulated to describe exotic baryon $\Theta^+$, to dibaryons .

## 4 Dibaryon states

In 1977 a bound six-quark state ($uuddss$),the H -dibaryon,was predicted in a bag-model calculation by Jaffe[18].This state is the lowest $SU_f(3)$ flavor singlet state with spin zero, strangeness -2 and $J^P = 0^+$

We consider the direct product space of the $J^\pi = \frac{1}{2}^+$ baryon octet with itself in terms of irreducible representations of $SU_f(3)$

$$\mathbf{8} \otimes \mathbf{8} = \mathbf{1} \oplus \mathbf{8} \oplus \mathbf{8} \oplus \mathbf{10} \oplus \overline{\mathbf{10}} \oplus \mathbf{27}.$$

The hypercharge (Y) ranges from +2 (NN,nucleon-nucleon) to -2 ( $\Xi\Xi$ ) for these dibaryon states. The Y = 2 member include the deuteron of the $\overline{10}$ - multiplet. The Y = 1 states contain $N\Lambda$ and $N\Sigma$ and the Y= 0 members contain $\Lambda\Lambda$ , $N\Xi$ , $\Lambda\Sigma$ and $\Sigma\Sigma$ dibaryons. There is several doubly-strange dibaryon states in the $SU_f(3)$ multiplets which do not directly decay to $\Lambda\Lambda$ and may appear as resonances in the $\Lambda\Lambda$ and $N\Xi$ systems .The lowest - order processes in $K^- d \to K^0 \Lambda\Lambda$ reaction resulted to H - dibaryon production is shown in fig.4.

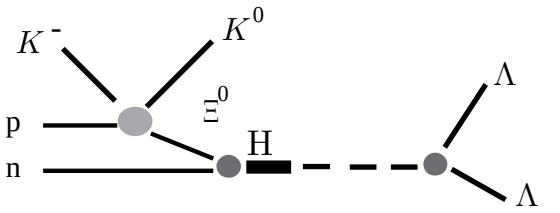

FIG.4: The lowest-order processes in $K^- d \to K^0 \Lambda\Lambda$ reaction.

One of the great open problems of intermediate energy physics is the question of existence or nonexistence of dibaryons. Early theoretical models based on SU(3) and SU(6) symmetries [19,20] and on Regge theory [19,21] suggest that dibaryons should exist. There is QCD-based models predict dibaryons with strangeness S= 0, -1, and -2. The invariant masses range between 2 and 3 GeV [18,22,23,24,25,26,27,28,29,30] . The masses and widths of the expected 6-quark states differ considerably for these models. But it seems that all QCD models predict dibaryons and none forbids them. Until now, about 30 years after the first predictions of the S=-2 H-dibaryon by Jaffe this question is still open.

We consider H dibaryon, fig.5, which composed of uuddss, three vector ud - ud - ss diquarks, each of which is a colour antitriplet and is symmetric in flavour and spin and orbital space, this leads to a six - quark state which is colour singlet, we ignore the pauli principle for quarks in different diquarks in the limit that diquarks are pointlike, but two quarks in each diquark satisfy this principle.

Now by using of diquark ideas in the chiral limit diquark correlations in the relativistic region and imposing HF interactions between quarks in a diquark, we led to introducing a conventional Hamiltonian

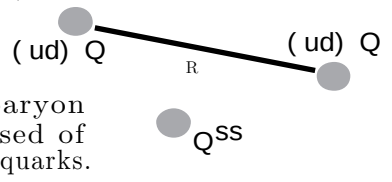

FIG.5 :H dibaryon which composed of three vector diquarks.

$$H(H) = T(H) + V^{cs}(H) \qquad (19)$$

with

$$T(H) = \frac{\vec{\nabla}_R^2}{2m}$$

The orbital wave function is

$$\psi_m = N[a_R Y_{1m}(\hat{R}) e^{-a^2 R^2/2}] \qquad (20)$$

We calculated the masses of vector diquarks using colour - spin interactions as HF intraction between quarks in a diquark.

If we take over the results and consider a T=450 Mev Kinetic energy as the binding energy for the H dibaryon the mass of it would be equal to the sum of the $\Xi^0$ and N masses.

We suppose that Decay widths of H is due to tunneling of one of the quarks between the two vector diquarks.

In the decay process $H \to \Xi^0$ N a (d) quark tunnels from a diquark ud to the other diquark to form a nucleon (udd) and an off-shell (u) quark which forms $\Xi^0$ with the other diquark .

According to our model the Kinetic energy $T \simeq \frac{\vec{\nabla}_R^2}{2m} = \frac{3a}{4m} \simeq 450$MeV this leads to $a \simeq 444$ and then

$$r_0 = \langle R \rangle = \sqrt{\frac{5}{2a^2}} \simeq 0.003 \text{ MeV}^{-1} \qquad (21)$$

we have $g^2 = 3/03$, $g_A = 0.75$ from the quark model and $\Delta E = (m_u + m_d) - M_{ud}$ where $M_{ud} \simeq 520$ MeV in our model.

Inserting this values into Eq (15) we find

$$\Gamma_H \simeq 45 \text{ MeV}. \qquad (22)$$

Now the paper consider another six-quark state uu-dd-cc ($H_{cc}$) composed of three vector diquarks.The decay width of the process $H_{cc} \to X_U N$ would be

$$\Gamma_{H_{cc}} \simeq 52 \text{ MeV}. \qquad (23)$$

Another method for calculating the mass and stability of the $\Theta^+$ pentaquark is based on our diquark - diquark - antiquark approach of our model in which we have used a long - range nonperturbative binding energy for the confinement interaction between two diquarks ( QQ ) ,fig.6, in the form

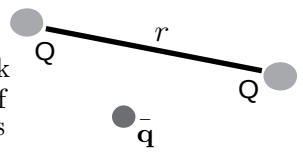

FIG.6 : $\Theta^+$ pentaquark which composed of two vector diquarks and one antiquark.

$$V(r_i, r_j) = a\,|\vec{r}_i - \vec{r}_j|^2 \qquad (24)$$

$$r = |\vec{r}_i - \vec{r}_j| \qquad (25)$$

where $a$ is the binding streanth.

In the model the two light quarks compose a bound diquark system in the antitriplet colour state with HF ( CS , FS ) interactions between quark pairs; now we suppose each diquark as a localized colour source in which the light quarks moves; interaction forces between diquarks ( Q's ) then lead to the formation of a two - particle bound state of a QQ system, the scale of which is determined by the quantity 1 / $m_Q$ which is smaller than the QCD scale 1 / $\Lambda_{QCD}$ ( $\Lambda_{QCD}$ is 200 ~ 400 MeV ).

We assume the light antiquark influence on the diquark ( Q ) dynamics is small due to universal nature of diquarks. For the $\Theta^+(1540)$, particle production is not kinematically possible and the motion is non-relativistic ( $v^2/c^2 = 0.08$ for the nucleon and 0.30 for the kaon ).Therefore it should be possible to use the methods of non-relativistic theory based on schröinger equation.

Thus we use a shrödinger - like equation for the QQ system as

$$H\,\psi(r) = E\,\psi(r) \qquad (26)$$

where

$$H = T + V(r) \qquad (27)$$

and E is the total binding energy of the system, this leads to

$$E = a/m \qquad (28)$$

if we consider $\psi(r) = N e^{-ar^2}$ for the wave function of the system.

The expectation value of potential is

$$\langle V \rangle = 3a/4a. \qquad (29)$$

Now for the second version of our model we have $M_{ud} \simeq$ 520Mev and $m_s \simeq 450$ MeV, thus if we consider E $\simeq$ 50 MeV, the resulted mass of $\Theta^+$ pentaquark would be 1540 Mev .

Finding (a) from Eq (28) and inserting it into

$$\langle r^2 \rangle = 3/4a \qquad (30)$$

leads to

$$\bar{r}_0 \simeq 0.009 \text{ MeV}^{-1} \qquad (31)$$

for the relative distance of diquarks in the QQ system.

This result is in agreement with our previous one Eq (17) and leads to $\Gamma \simeq 1.30$ MeV width for $\Theta^+$ pentaquark using of tunneling approach.

## 5 Conclusions

Assuming a (CP$\chi$QM) effective field theory, we suggested that the $\Theta^+$ baryon is a bound state of two vector diquarks and a single antiquark, the spatially wave function of these diquarks has a P-wave and a S -wave in angular momentum in the first and second version of our model respectively.

At the first we constructed the total OCFS symmetry of $q^4\bar{q}$ contri bution of quarks.

Then by using of diquark ideas in the chiral limit diquark correlations in the relativistic region and imposing HF interactions between quarks in a diquark, we led to introducing a conventional Hamiltonian and by considering its solution for the second version of our model we led to $\Theta^+$ pentaquark mass.

According to these considerations the paper calculated the average distance between two diquarks which leads to a reasonable width $\Gamma_{\Theta^+} \simeq 1.30$ MeV for the $\Theta^+$ due to tunneling model. The observed width of $\Theta^+$ is unexpectedly narrow, considering that the decay of $\Theta^+$ requires no pair creation and goes through a fall-apart process.[31] Attempts have been made to explain the width based on the quark model[32] and QCD sum rule.[33-35] However most of them suggest the symmetry properties of the orbital - colour spin - flavour wave function. Furthermore, such choises of the wave function seem quite artificial.

Although the 2005 JLab experiment [36] found no pentaquarks in the mass region 1480 MeV $< m_{\Theta^+} <$ 1700 MeV, our paper predicts that for $m_{\Theta^+} >$1700 MeV its width $\Gamma_{\Theta^+} >$ 53 MeV, and there is a possibility for the existence of pentaquark states which are too wide to be detected by experiments. the confinement[37] potential between two vector diquarks, which the distance between them $r_0 < 0.005$ , may be responsible for the stability of such multiquark states.

Now there is new strong evidence of an extremely narrow $\Theta^+$ from DIANA,[17] in reaction $K^+n \to K^0p$ the mass of the $pK^0$ resonance is $m = 1537 \pm 2$ MeV/c$^2$ with an upper limit on intrinsic width $\Gamma = 0.36 \pm 0.11$ MeV/c$^2$ which is in agreement with our width limit.The null results from CLAS are compatible with expectations based on cross section considerations.[38,39]

By extendeding our diquark model to dibaryons the paper calculated some estimations of the mass and width of the H dibaryon.

In an another method by using of a shrödinger - like equation for the QQ system in our model, we calculated the mass and the width of the $\Theta^+$ pentaquark. Our results in this method, for example the average distance between diquarks, are in agreement with our previous conclusions.

Our theoretical results on the mass and width of $\Theta^+$and H are in agreement with many experimental limits and one can use the vector diquark approach for simultaneously calculatig the mass and width of other multiquark states, for example those multiplet states which we have introduced in Eq(2) for pentaquarks in which they have vector diquarks.

## Acknowledgment

I would like thank Arifa Ali Khan for numerous discussions and reading the manuscript and for very useful comments.

## References

[1] Particle Data Group, S. Eidelman *et al.*,Phys. Lett. B **592**, 1 (2004).
[2] A. Cabo Montes de Oca *et al.*, Eur. Phys. J. C **47**, 355, (2006).hep-ph/0008003 v4 22 May (2002).
[3] A. R. Haghpayma, hep-ph/0606214 v1 20 Jun (2006), hep-ph/0606270 v1 26 Jun (2006).
[4] M. Ida, R. Kobayashi, Progr. Theor. Phys. **36**, 846 (1966).
[5] D.B.Lichtenberg, L. J. Tassie, Phys. Rev. **155** , 1601 (1967).
[6] M. Kirchbach, M. Moshinsky, Y. F. Smirnov, Phys. Rev.D **64**,114005 (2001).
F. Wilczek, hep-ph/0409168 v2 17 Sep (2004),R. L. Jaffe, Phys. Rept. 409, 1(2005).
[7] M. Anselmino *et al.*, Rev. Mod. Phys. **65**, 1199 (1993). W. Brodowski *et al.*,Z. Phys. A 355, 5 (1996).
[8] D. Ebert *et al.*, Phys.Rev.C **72** , 015201 (2005), hep-ph/0608304 v2 17 Nov (2006).
J. Bardeen *et al.*,Phys. Rev. 108 (5), 1175 (1957).
[9] O. Jahn *et al.*,PoS LAT, 069 (2005), hep-lat/0509102 v1 23 Sep (2005).
[10] M. Oka, Int. J. Mod. Phys.A**21**, 807 (2006), hep-ph/0509060 v1 7 Sep (2005).
[11] V. Bernard and U-G. Meissner, commissioned article for Ann. Rev. Nucl. Part. Sci, hep-ph/0611231.
[12] A. Ali Khan, hep-lat/0507031 v3 24 Oct (2005).
[13] R. Jaffe, F. Wilczek, Phys. Rev. Lett. 91 , 232003 (2003).
[14] T. Inoue *et al.*, Int. J. Mod. Phys. E**14** ,995 (2005), hep-ph/0407305 v2 25 Nov (2005).
[15] Deog Ki Hong *et al.*, Phys. Lett. B**596** ,191 (2004), hep-ph/0403205 v2 3 Jan (2004).
[16] Volker D. Burkert, Int. J. Mod.Phys.A **21**,1764 (2006), hep-ph/0510309 v2 7 Nov (2005).
A. G. Oganesian, hep-ph/0608031 v1 3 Aug (2006).
[17] V. V. Barmin *et al.* [DIANACollaboration], hep-ex/0603017 v2 21 Apr (2006).
[18] R.L. Jaffe, Phys. Rev. Lett.**38** , 195 (1977).
[19] R. J. Oakes, Phys. Rev .**131** ,2239 (1963).
[20] F. J. Dyson, N.H. Xuong,Phys. Rev. Lett.**13** ,815 (1964).
[21] Fl. Stancu *et al.*,Phys. Rev.D **57**, 4393 (1999).
[22] S. Graffi, Lett. Nuo. Cim. 2, 311 (1969).
[23] A.T. M. Aerts *et al*, Phys. Rev. D **17** (1978) 260;D **21** , 1370 (1980).
[24] P. J. G. Mulders *et al.*, Phys. Rev.D **19**, 2635 (1979);D **21**, 2653 (1980).
[25] R. L. Jaffe *et al.*, Nucl. Phys. A **625** , 167 (1997), hep-ph/9705407 v1 23 May (1997).
Tsutomu Sakai *et al.*, Phys. Lett. B **430** , 168 (1998), hep-ph/9708433 v2 7 May (1998).
[26] C. W. Wong, Prog. Part. Nucl. Phys. 8, 223 (1982).
[27] Glennys R. Farrar *et al*, Phys. Rev. D**70** , 014008 (2004), hep-ph/0308137 v1 12 Aug (2003).
Shi-Lin Zhu, Phys. Rev. C **70** , 045201 (2004), hep-ph/0405149 v2 18 May (2004).
Hourong Pang *et al.*, Phys. Rev. C **69** , 065207 (2004), nucl-th/0306043 v2 2 Apr (2004).
[28] A. T. M. Aerts, C. B. Dover, Nucl. Phys. B **253** ,116 (1985).
[29] P. La France, E. L. Lomon, Phys. Rev. D **34**, 1341 (1986), P. Gonz lez, E. L. Lomon, ibid. D34, 1351 (1986), P. Gonz lez *et al.*, ibid. D **35** ,2142 (1987).
[30] Yu. S. Kalashnikova *et al.*,Yad. Fiz. 46, 1181 (1987) ,trans. Sov. J. Nucl. Phys. **46**, 689 (1987).
N. Konno *et al.*, Phys. Rev. D **35**, 239 (1987),T. Goldman *et al.*, Phys. Rev. C 39, 1889 (1990).
[31] R. L. Jaffe, A. Jain, Phys. Rev. D **71**, 034012 (2005),hep-ph/0408046 v2 28 Jan (2005).
[32] D. Melikhov *et al.*, Phys. Lett,B **594** 265 (2004), hep-ph/0405037 v2 27 May (2004).
[33] M. Eidemuller *et al.*,Phys. Rev. D **72** , 034003,(2005),hep-ph/0503193 v1 18 Mar (2005).
[34] Z. G. Wang *et al.*, Phys. Rev.D **72**, 034012 (2005),hep-ph/0504151 v3 31 Jul (2005).
[35] D. Melikhov, B. Stech, Phys. Lett. B**608**, 59 (2005),hep-ph/0409015 v2 8 Des (2004).
[36] CLAS collaboration, B. Mckinnon *et al.*, Phys. Rev. Lett. 96, 212001 (2006) ,hep-ex/0603028 v1 14 Mar (2006),R. De Vita *et al.*[CLAS Collaboration] , Phys. Rev. D **74**, 032001 (2006), hep-ex/0606062 v1 27 Jun (2006), S. Niccolai *et al.* [CLASCollaboration] ,Phys. Rev. Lett. **97**, 032001 (2006).
[37] R. Alkofer *et al.*, hep-ph/0610365 v1 27 Oct (2006).
[38] Dmitri Diakonov., hep-ph/0610166 v1 13 Oct (2006).
[39] V. Guzey., hep-ph/0608129 v1 11 Aug (2006).